# Trapped field and related properties in a superconducting disk magnetized by pulse field

**C Y He, Y Hou, L Liu, Z X Gao**

Department of Physics, Key Laboratory for Artificial Microstructure and Mesoscopic Physics, Peking University, Beijing 100871, P. R. China

**Abstract**

The distributions of the magnetic field and temperature in a superconducting-disk magnetized by pulse field, and the levitation force between this disk and a permanent magnet are calculated from first principles. The calculation is based upon the current motion and the heat diffusion equations in the disk. The critical current density as a function of magnetic field and temperature is taken into account. The dissipation power in the superconducting or the normal state region is distinguished. The trapped field may reach 17 T at 29 K by pulsed field magnetization. The effects of the amplitude of the current pulse on the distributions of magnetic field in the disk and on the levitation force are discussed.

## 1. Introduction

With the rapid development in fabrication of large sized high-temperature-superconductors (HTS), HTS have attracted intense attention for their strong pinning effect and great ability to trap high magnetic fields. Recently it is reported that a melt texture growth Y-Ba-Cu-O disk of 2.65 cm diameter can trap fields over 17 T at 29 K [1]. Therefore, HTS are believed to be promising to act as permanent magnets that are superior to conventional magnets from the point of view of practical applications [2]. Many different potential engineering applications have emerged for such magnets [3-9], and some have even been commercialized [5-7]. Magnetic levitation trains, flywheels and magnetic bearings have been developed by utilizing the repulsive force against the magnetic fields produced by a permanent magnet. The HTS with trapped magnetic fields are used as permanent magnets in superconducting motors and magnetic field generators capable of producing fields higher than 2 T [10].

Three typical ways are often used to magnetically activate HTS: field-cooling magnetization (FC), zero-field-cooling magnetization (ZFC) and pulsed field magnetization (PFM) [11]. FC, which is usually used in the laboratory, is the most effective method to extract the potential of the material. However, an intense static field to magnetize HTS requires large-scale conventional superconducting solenoid magnets; while what PFM needs is only a small and simple coil to apply a magnetic field, which makes it possible to design a more compact and convenient electromagnetic system with an *in situ* PFM coil [11]. In this paper we will only discuss a superconducting-disk (SD) magnetized by PFM.

The induced current density and the magnetic field in the SD are the key properties for its applications. We have already reported a calculation of the levitation force between a permanent magnet and a SD magnetized in an axial applied uniform magnetic field before [12], which omitted the heat dissipation in the SD. H. Ohsaki *et al*. have reported the pulsed field magnetization by taking into account the critical current density as a function of magnetic field and temperature and the heat dissipation, and given the distributions of current density and temperature [11]. However, it is not clear whether or not they distinguish the normal state region from the superconducting state region both in the calculations of current density and heat dissipation. In this paper we have managed to solve this problem by taking different material equations $E(J)$. This distinction is very important especially when we consider the heat dissipation. In their work the maximum trapped magnetic field is only 0.25 T at the end of PFM, and the long-time trapped magnetic field will be much smaller than the value of a permanent magnet, such as one made of Nd-Fe-B. We try to obtain higher trapped magnetic field and prefer to choose a co-axial PFM coil instead of two pulsed coils for convenience in practical magnetic levitation applications. After all, we have calculated the distribution of the magnetic field and the temperature in the SD magnetized by PFM, and the levitation force between the SD and a permanent magnet from first principles. Numerical analysis is used in these calculations. How the levitation force depends on the amplitude of the pulse current is also carefully discussed. The levitation force with the heat dissipation in the SD magnetized by superconducting magnet taken into account will be discussed elsewhere.

## 2. Modeling

### A. Configuration

We consider a SD with diameter $2a$ and thickness $2b$, inset in a co-axial magnetizing coil with outer diameter $6a+0.4a$ and thickness $2b$. In order to keep the temperature of the sample surface fixed, the inner diameter is taken as $2a+0.4a$ in order to leave enough places for cooling medium to contact the SD closely. The center of the SD is taken as the origin of the cylindrical coordinate system $(r, \phi, z)$.

Now we consider the vector potential of one circular current loop [13], as shown in Fig. 1,

$$A_\phi^1(r, a, z) = \frac{\mu_0 I_p}{4\pi} \oint \frac{ds_\phi}{r} = \frac{\mu_0 I_p}{2\pi} \int_0^\pi \frac{a \cos\phi d\phi}{(a^2 + r^2 + z^2 - 2ar\cos\phi)^{1/2}} \tag{1}$$

where $I_p$ is the current in the circular loop. We define

$$f(r, r', z-z') = \int_0^\pi \frac{d\phi}{2\pi} \frac{-r'\cos\phi}{\left[(z-z')^2 + r^2 + r'^2 - 2rr'\cos\phi\right]^{1/2}}$$

$$= \frac{-1}{\pi k} \sqrt{\frac{r'}{r}} \left[\left(1 - \frac{1}{2}k^2\right)K_1(k^2) - K_2(k^2)\right], \tag{2}$$

where

$$k^2 = \frac{4rr'}{(r+r')^2 + (z-z')^2}. \tag{3}$$

$K_1$ and $K_2$ are the complete elliptic integrals of the first and second kind respectively, and then $A_\phi$ can be written as

$$A_\phi^1(r, r', z) = -\mu_0 I_p f(r, r', z) \tag{4}$$

For the magnetizing coil specified above, we can get its vector potential $A_\phi$ by integrating $A_\phi^1$:

$$A_\phi(r, z) = -\mu_0 \int_{a+0.2a}^{3a+0.2a} dr' \int_{-b}^{b} dz' j_p f(r, r', z-z') \tag{5}$$

where $j_p$ is the current density in the coil. Using the integral kernel with $\mathbf{r} = (r, z)$, $\mathbf{r'} = (r', z')$

$$Q_{cyl}(\mathbf{r}, \mathbf{r'}) = f(r, r', z-z'), \tag{6}$$

it reads

$$A_\phi(r, z) = -\mu_0 \int_{a+0.2a}^{3a+0.2a} dr' \int_{-b}^{b} dz' j_p Q_{cyl}(\mathbf{r}, \mathbf{r'}) \tag{7}$$

**B. Basic Equation**

Using the Maxwell Equation,

$$\nabla \times \vec{B} = \boldsymbol{\mu}_0 \vec{J} \tag{8}$$

we can determine the profile of the magnetic field $\vec{B}$, the electrical field $\vec{E}$ and the current density $\vec{J}$ in the SD with the material law by following the same procedures in our former paper [12]. We finally get the basic current motion equation,

$$\dot{J}(\mathbf{r},t) = \boldsymbol{\mu}_0^{-1} \int_0^a d\mathbf{r}' \int_{-b}^b dz \, \mathcal{Q}_{cyl}^{-1}(\mathbf{r},\mathbf{r}') \left[ E(J) + \dot{A}_F(\mathbf{r}',z') \right] \tag{9}$$

where $A_F$ is the vector potential of the applied magnetic field and $Q_{cyl}(\mathbf{r},\mathbf{r}')$ has the same meaning defined in part A.

$E(J) = E_c \left( \dfrac{J}{J_c} \right)^n$ for the superconduting state region ($J < J_c$), $s = n-1$ is the vortex creep exponent, and

$E(J) = E_c \left( \dfrac{J}{J_c} \right)$ for the normal state region ($J \geq J_c$). Eq. (9) can be easily time integrated by starting with

$J(\mathbf{r},z,t_0) = J_0$ ($J_0$ is the initial current density distribution in the SD) and then by putting

$J(\mathbf{r},z,t = t+dt) = J(\mathbf{r},z,t) + \dot{J}(\mathbf{r},z,t)dt$.

As soon as the induced current density $J(\mathbf{r},z,t)$ is obtained, the vector potential generated by the induced current density $A_J$ can be derived as,

$$A_J(\mathbf{r},z) = -\boldsymbol{\mu}_0 \int_0^a d\mathbf{r}' \int_{-b}^b dz' Q(\mathbf{r},\mathbf{r}') J(\mathbf{r}'). \tag{10}$$

And the radial and axial trapped magnetic field can be written in the form of,

$$B_r = -\frac{\partial A_J}{\partial z} \quad , \quad B_z = \frac{1}{\boldsymbol{r}} \frac{\partial (\boldsymbol{r} A_J)}{\partial \boldsymbol{r}} \tag{11}$$

respectively.

**C. The permanent magnet (PM)**

When the distributions of the current density and the magnetic field are relatively stable in the SD, for example at $t = 100 \, \text{s}$ after PFM, the SD is levitated over a co-axial cylindrical PM with radius $R_{PM}$ and thickness $d_{PM}$. According to the similar discussion as before [12], the top surface center of the PM approaches and recedes from the

SD as,

$$s = z_0 + z_0 \cos(\boldsymbol{w} \cdot t) + b + z_{00} \tag{12}$$

where frequency $\boldsymbol{w}$ represents the speed at which the PM approaches and recedes from the SD, $z_{00} + 2z_0$ is the initial distance and $z_{00}$ is the minimum distance between the top surface of the PM and the bottom surface of the SD. We choose $z_{00}/z_0 = 0.1$ to avoid uncertainty in this calculation [12].

We finally get the vector potential of the PM along the $\boldsymbol{f}$ direction [12],

$$A_f(\boldsymbol{r}, z) = \frac{B_{\text{rem}}}{2\boldsymbol{p}} \int_0^{\boldsymbol{p}} R_{\text{PM}} \cos \boldsymbol{f} \ln \frac{(z + s + d_{\text{PM}}) + \sqrt{R_{\text{PM}}^2 + \boldsymbol{r}^2 - 2\boldsymbol{r}R_{\text{PM}} \cos \boldsymbol{f} + (z + s + d_{\text{PM}})^2}}{z + s + \sqrt{R_{\text{PM}}^2 + \boldsymbol{r}^2 - 2\boldsymbol{r}R_{\text{PM}} \cos \boldsymbol{f} + (z + s)^2}} d\boldsymbol{f} \tag{13}$$

where $B_{\text{rem}}$ is the remnant induction of the PM. Then the radial field $B_r^{PM} = -\partial A_f / \partial z$ can be written as,

$$B_r^{PM}(\boldsymbol{r}, z) = \frac{B_{\text{rem}}}{\boldsymbol{p}} \sqrt{\frac{R_{\text{PM}}}{\boldsymbol{r}}} \sum_{i=0}^{1} \frac{(-1)^i}{k_i} \left[ \left(1 - \frac{1}{2}k_i^2\right) K_1\left(k_i^2\right) - K_2\left(k_i^2\right) \right] \tag{14}$$

where $K_1$ and $K_2$ are the complete elliptic integrals of the first and second kind respectively. And

$$k_i^2 = \frac{4\boldsymbol{r}R_{\text{PM}}}{(R_{\text{PM}} + \boldsymbol{r})^2 + (z + s + id_{\text{PM}})^2}, \quad i = 0, \ 1 \tag{15}$$

From the equations listed in this part and Eq. (9), we can calculate the trapped magnetic field and the induced current density in the SD during all stages of the process in which the PM approaches and recedes the SD. When we solve Eq. (9) in this process, the current density at $t = 100 \text{ s}$ after PFM is used as the initial current density, and we should also notice that $A_f(\boldsymbol{r}, z)$ here is generated by the PM instead of by PFM.

**D. Heat dissipation**

In order to get the temperature distribution in the SD, we should take the heat dissipation in the SD into account. The heat diffusion equation is given as,

$$C\partial T / \partial t - \boldsymbol{k}\nabla^2 T = W \tag{16}$$

where $\boldsymbol{k}$ is the thermal conductivity, $C$ is the heat capacity per volume, and $W = \bar{E} \cdot \bar{J}$ is the heat power generated in the SD per volume. This partial differential equation can be solved numerically. When we solve the equation, appropriate boundary conditions should be chosen.

In our calculation, we use an adiabatic boundary condition $\frac{\partial T(\vec{r},t)}{\partial n}\big|_{\Sigma}=0$ during PFM [11], which assumes that this process is too short for the SD to exchange heat with the surroundings. We take $T_0$ as the uniform initial temperature. The solution to Eq. (14) in these conditions can be expressed as,

$$T(\vec{r},t)=T_0+\int_0^t dt'\iiint\limits_{V'} W(\vec{r}',t')G(\vec{r},t;\vec{r}',t')d\vec{r}' \tag{17}$$

where the Green's function $G(\vec{r},t;\vec{r}',t')$ satisfying the boundary conditions $\frac{\partial G(\vec{r},t;\vec{r}',t')}{\partial n}\big|_{\Sigma}=0$ is given by,

$$G(\vec{r},t;\vec{r}',t')=T_0(t)+\sum_{m=1}^{\infty}T_{0m}J_0(\frac{m_0^m}{a}r)+\sum_{n=1}^{\infty}T_{0m}\cos(\frac{n\mathbf{p}}{b}z)+\sum_{m=1}^{\infty}\sum_{n=1}^{\infty}T_{mn}(t)J_0(\frac{m_0^m}{a}r)\sin\frac{n\mathbf{p}}{b}z \tag{18}$$

Here $m_n^m$ is the $m$th positive root of Bessel function $J_n(x)$.

In the following time after PFM, however, we choose a fixed temperature boundary condition $T(\vec{r},t)\big|_{\Sigma}=T_0$, which assumes ideal thermal contact with the surroundings at the temperature $T_0$. The temperature profile at the end of PFM is taken as the initial temperature in this process. The solution to Eq. (16) in these conditions then can also be written in the form of Eq. (17). But in this case the Green's function satisfying $G(\vec{r},t;\vec{r}',t')\big|_{\Sigma}=0$ has a different form as,

$$G(\vec{r},t;\vec{r}',t')=\sum_{m=1}^{\infty}\sum_{n=1}^{\infty}\frac{1}{C}\frac{\frac{1}{2\mathbf{p}}J_0(\frac{m_0^m}{a}r')\sin\frac{n\mathbf{p}}{b}z'}{\frac{a^2}{2}[J_0(m_0^m)]^2\frac{b}{2}}e^{-\frac{k}{C}[(\frac{n\mathbf{p}}{b})^2+\frac{m_0^m}{a})^2](t-t')}\mathbf{h}(t-t')J_0(\frac{m_0^m}{a}r)\sin\frac{n\mathbf{p}}{b}z \ . \tag{19}$$

The heat dissipation is taken into account by $J_c$ as a function of the temperature and the magnetic field, which is a constant for simplicity in our former work [12]. In our calculation the dependence of $J_c$ on the magnetic field can be described by the Kim model [11],

$$J_c=J_{c0}\frac{B_0}{|B|+B_0} \tag{20}$$

where $J_{c0}$ is $J_c$ at $B=0$ and $B_0$ is a constant. The temperature dependence of $J_{c0}$ is given by [11]

$$J_{c0}(T)=\mathbf{a}\left\{1-\left(\frac{T}{T_{c0}}\right)^2\right\}^2 \tag{21}$$

where $T_{c0}$ is the critical temperature of the SD for $B=0$ and $\mathbf{a}$ is a constant.

From the above discussions, we can obviously see that Eq. (9) and the series of Eq. (16), (20) and (21) are coupled from each other. In order to get the numerical solution, we can deal with these equations in the following steps, which are similar to the procedure reported by S. Braeck *et al*. [12]. First, $J_c$ is taken as the value at the initial temperature and magnetic field, then $J(\bar{r},t)$, $E(\bar{r},t)$ and $B(\bar{r},t)$ with the time evolution can be calculated by Eq. (9). Second, $W(\bar{r},t)$ in Eq. (16) is substituted with the calculated $J(\bar{r},t)$ and $E(\bar{r},t)$. We must emphasize $E = E_c \left( \dfrac{J}{J_c} \right)^n$ for $J < J_c$ and $E = E_c \left( \dfrac{J}{J_c} \right)$ for $J > J_c$ in order to distinguish the superconducting or normal state region. Then the temperature distribution $T(\bar{r},t)$ can be deduced with Eq. (17) and the corresponding Green function expression. In the next step, we can recalculate the temperature and the magnetic field dependent $J_c(\bar{r},t)$ according to Eq. (20) and Eq. (21) with $B(\bar{r},t)$ and $T(\bar{r},t)$ attained in the first step. Finally, we can educe the corrected $J(\bar{r},t)$ and $B(\bar{r},t)$ with new $J_c(\bar{r},t)$ via Eq. (9). We should repeat these procedures until a self-consistent numerical solution is obtained, which means the solution can be thought the same with the result with more repetitions in a certain tolerance.

## E. The levitation force

As the current density $J(\mathbf{r},z,t)$ and the radial magnetic field $B_r^{PM}$ inside the SD have been derived, the vertical levitation force along the *z*-axis can be readily obtained as [12],

$$F_z = 2\boldsymbol{p} \int_0^a \boldsymbol{r} \, d\boldsymbol{r} \int_{-b}^b dz J(\mathbf{r},z) B_r^{PM}(\mathbf{r},z) \qquad (22)$$

## 3. Results and discussions

## A. Selection of parameters

The parameters used in our calculation are taken as follows: $2a = 2.65\,\mathrm{cm}$, $2b = 1.5\,\mathrm{cm}$ [1]; $T_0 = 29\,\mathrm{K}$ [1]; $E_c = 1\times10^{-4}\,\mathrm{V/m}$ [11]; $B_0 = 0.4\,\mathrm{T}$ [11]; $T_{c0} = 92\,\mathrm{K}$ [11]; $\boldsymbol{s} = 20$ [12]; $R_{PM}/a = 1$ [12]; $d_{PM}/b = 1$ [12]; $z_0/a = 0.5$ [12]; $z_{c0}/z_0 = 0.1$ [12], $B_{rem} = 0.5T$ [12]; $\boldsymbol{w} = 0.1$ [12]; $\boldsymbol{m}_0 = 4\boldsymbol{p} \times 10^{-7}$ ; $C = 0.88\times10^6\,\mathrm{J/m^3 K}$ [14]; $\boldsymbol{k} = 6\,\mathrm{W \cdot m^{-1} \cdot K^{-1}}$ [14]. In order to get $J_c(29\,\mathrm{K}, 0\,\mathrm{T}) = 1.0\times10^9\,\mathrm{A/m^2}$, we take $\alpha = 1.23 \times 10^9 \mathrm{A/m^2}$. All these parameters remain unchanged in the following discussions unless special declaration.

## B. Distribution of the magnetic field in the SD

We can discuss the time evolution of the flux density profile during PFM. The form of the pulse current is $j_p = j_1 \sin\dfrac{2\boldsymbol{p}t}{t}$ with the pulse width $t/2 = 6\,\text{ms}$ [11]. The value of $j_1$ may be taken in the range of $1\sim9\,j_{10}\,\left(j_{10}=1.0\times10^9\ \text{A/m}^2\right)$ [15]. The time evolution of the magnetic field during PFM in the SD at $T = 29$ K for $j_1 = 1.0\,j_{10}$ is shown in Fig.2. The magnetic field includes two parts. One is the applied magnetic field generated by PFM, and the other is generated by the induced current density in the SD. The magnetic field is first induced in the peripheral region of the SD. Then it will spread into its inner region. When the pulse current in the coil rises to its top, the applied magnetic field reaches its maximum. Afterwards the applied magnetic field decreases thus the induced magnetic field changes its direction and the maximum magnetic field in the SD will extend towards the disk center.

The distributions of trapped magnetic field in the SD for different $j_1$ [=1, 1.5, 2, 2.5, 3 $j_{10}$ ] at $t = 100$ s are shown in Fig. 3(a). With the increase of $j_1$ the maximum trapped magnetic field increases, the two peaks of the trapped magnetic field will be close to each other and finally become one platform in the center of the SD. It is clear that the maximum trapped magnetic field increases with the pulsed current magnitude, but is restricted by the critical current density of the SD. For $j_1 = 3.0\,j_{10}$ it will be over 6 T. When $j_1 > 3.0\,j_{10}$, the maximum and the distribution of the trapped magnetic field will not change any more. If the critical current density $J_c\left(29\ \text{K, 0 T}\right)$ of the SD reaches $5\times10^9\ \text{A/m}^2$, the maximum trapped magnetic field may reach 17 T according to our calculation for $j_1 = 9.0\,j_{10}$ [15], shown in Fig. 3(b).

We have also calculated the distribution of the trapped magnetic field at $T = 77.4$ K. With the increase of the temperature the critical current density decreases and the flux motion is more active, so the trapped magnetic field will decrease at higher temperature. At $T = 77.4$ K for $J_c\left(77.4\ \text{K, 0 T}\right)=2\times10^8\ \text{A/m}^2$ the maximum trapped magnetic field may reach 1.2 T.

**C.  Distribution of the temperature in the SD**

The time evolution of the temperature in the SD for $j_1 = 1.0\,j_{10}$ is shown in Fig.4. During the first half pulse the temperature increase first emerges at the peripheral region of the SD, then spread to the center of the SD. At the end of the first half pulse, Fig. 4(a) shows that the temperature rise is the highest in the peripheral region, and continuously decreases to the minimum towards the center of the SD. After the first half pulse the applied magnetic field and the magnetic field at the peripheral region of the SD decrease, and the induced current will change its direction in the peripheral region of the SD. At the top of the magnetic field the induced current and the heat dissipation power is zero. The zero heat dissipation region spreads to the center of the SD with the top of the magnetic field, which results in the emergence of the minimum temperature increase region in the SD, shown in Fig.4 (b). At the end of the pulse the temperature increases almost in the whole SD. After PFM the temperature in the middle region of the SD rises a little and reaches its maximum value, and then decreases towards the circumstance temperature of 29K. For $k = 6 \ \mathrm{W \cdot m^{-1} \cdot K^{-1}}$ the maximum temperature is about 29.15 K in the middle region at about $t = 0.8 \ \mathrm{s}$, shown in Fig. 4(c).

There is no notable difference between the temperature evolutions during PFM with different $k$. It can be attributed to two factors. On one hand, the heat dissipation in the SD is the same for different $k$ under the adiabatic boundary condition. On the other hand, the pulse period of PFM is too short for different $k$ to produce any distinct difference. So the temperature distributions are almost the same with different $k$. After PFM, however, the thermal conductivity $k$ plays an important role in the temperature evolution. The back time for the whole SD to decrease to 29K is determined by the thermal conductivity $k$. For $k = 6 \ \mathrm{W \cdot m^{-1} \cdot K^{-1}}$ the back time is 17 s. The back time is shorter for larger $k$.

## D. Levitation force $F_z$

In practice the SD with the magnetizing coil is first cooled below $T_c$ at zero magnetic field and the current pulse is applied to the magnetizing coil at $t = 0$, and then at $t = 100 \ \mathrm{s}$ the PM is levitated over the SD. Fig. 5 shows the levitation force $F_z$ with different $j_1$ [=1, 1.5, 2, 2.5, 3 $j_{t0}$] during one cycle of harmonic motion of the PM. It is clear that the levitation force increases when the PM approaches the SD. And it is interesting that the hysteretic effect of the levitation force is too subtle to be observed. It can be attributed to the high induced current density by PFM, which is little influenced by the motion of the PM. So we can think that the induced current density by PFM dominates the levitation force.

It is obvious that the levitation force $F_z$ between the PM and the SD increase with $j_1$, which can be attributed to the increase of the induced current density and the radial magnetic field in the SD. However, we can not continuously enhance $F_z$ just by increasing $j_1$ due to the restriction of the critical current density thus the induced current of the SD.

The effects of the superconducting parameters, such as flux creep exponent $s$ and the critical current density $J_c$, on the magnetic levitation force by PFM are almost the same as that by the uniform field magnetization, which has been discussed before [12].

## 4. CONCLUSIONS

We have calculated the profile of the magnetic field and the temperature of a superconducting-disk magnetized by PFM from first principles. The vector potential generated by the magnetizing coil has been deduced on the basis of one circular current loop. The induced current density in the SD is obtained from the basic current motion equation. The heat dissipation in the SD is taken into account by considering the heat diffusion equation. We emphasize the heat dissipation in the superconducting state region or the normal state region must be distinguished. The amplitude of the current pulse plays an important role in the distributions of the trapped magnetic field and the levitation force according to our calculation. At $T = 29$ K for $J_c(0\,\text{T}) = 1 \times 10^9\,\text{A}/\text{m}^2$ the maximum trapped magnetic field is 6 T, and for $J_c(0\,\text{T}) = 5 \times 10^9\,\text{A}/\text{m}^2$ the maximum trapped magnetic field may reach 17 T. Due to the restriction of the critical current density of the SD, the trapped magnetic field and the magnetic levitation force can not continuously enhance by the increase of the magnitude of the pulsed current. The thermal conductivity $k$ has little influence on the temperature distribution during PFM, but plays an important role in the temperature distribution after PFM.

## ACKNOWLEDGMENTS

This work was supported by the National Science Foundation of China (NSFC 10174004) and the Ministry of

Figure captions:

Fig. 1: The coordinate of point P with respect to one circular current loop.

Fig. 2: The profile of the time evolution of the axial magnetic field on the top surface of the SD during PFM for $j_1 = 1.0\, j_{10}$.

Fig. 3: (a) The axial trapped magnetic field on the top surface of the SD for different $j_1$ [=1, 1.5, 2, 2.5, 3 $j_{10}$ ] at $t = 100$ s after PFM. (b) The axial trapped magnetic field on the top surface of the SD for $j_1 = 9.0 j_{10}$ at $J_c (29 \text{ K}, 0 \text{ T}) = 5 \times 10^9 \text{ A/m}^2$ .

Fig. 4: The distribution of the temperature in the SD for $j_1 = 1.0 j_{10}$ .

Fig.5: The levitation force $F_z$ for different $j_1$ [=1, 1.5, 2, 2.5, 3 $j_{10}$ ].

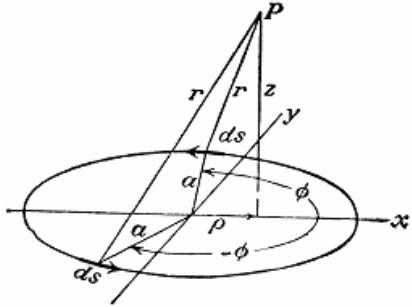

Fig.1

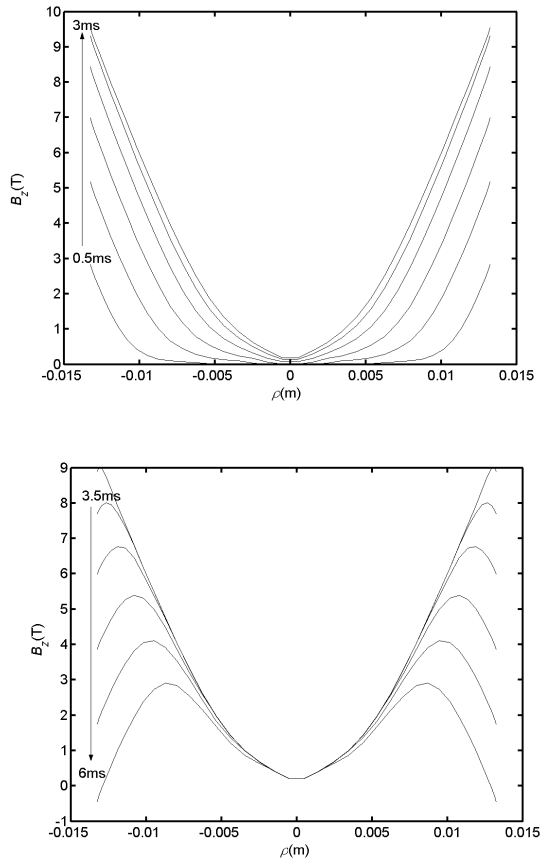

Fig. 2

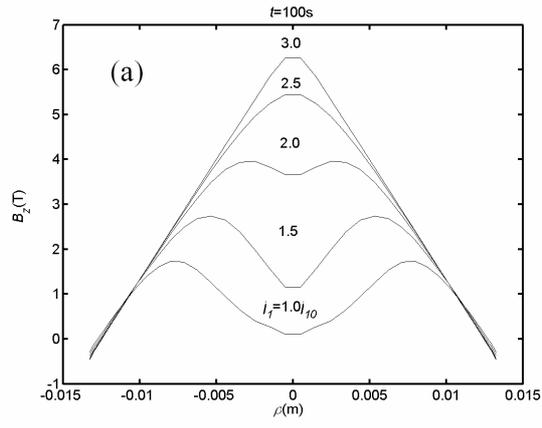

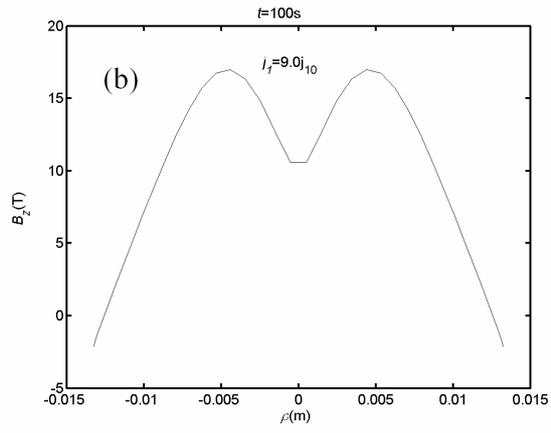

Fig. 3

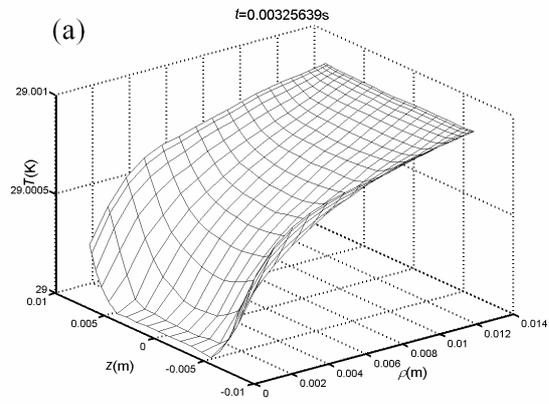

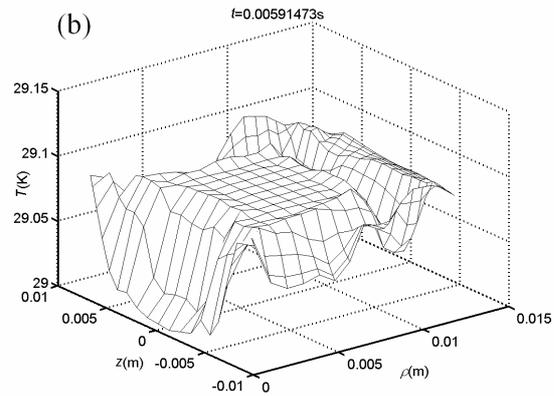

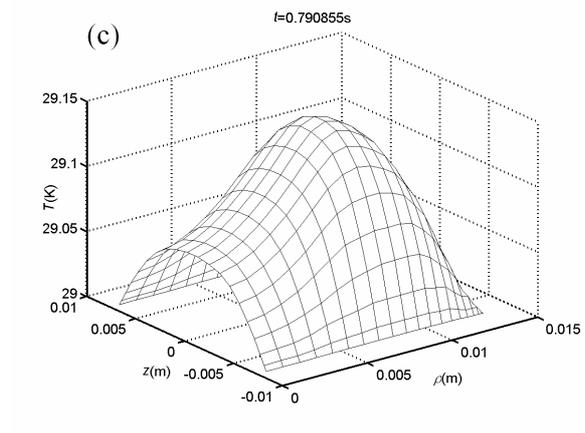

Fig. 4

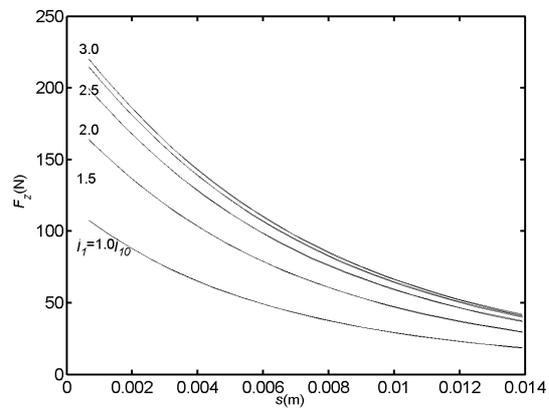

Fig.5